\documentclass[twocolumn,prb,citeautoscript,superscriptaddress,8pt]{revtex4-2}

\usepackage[hidelinks]{hyperref}
\usepackage{graphicx}
\usepackage{gensymb}
\usepackage{color}
\usepackage[dvipsnames]{xcolor}
\usepackage{float}
\usepackage{upgreek}
\usepackage{bm}
\usepackage{amsmath,amssymb}
\usepackage{multirow}

\begin{document}

\title{Stray magnetic field imaging of thin exfoliated iron halides flakes}

\author{Fernando Meneses} 
\thanks{These authors contributed equally to this work.}
\affiliation{School of Physics, University of Melbourne, Victoria 3010, Australia}
\affiliation{Centre for Quantum Computation and Communication Technology, School of Physics, University of Melbourne, Victoria 3010, Australia}

\author{Rongrong Qi} 
\thanks{These authors contributed equally to this work.}
\affiliation{National Graphene Institute, The University of Manchester, Manchester, UK}
\affiliation{Department of Physics and Astronomy, The University of Manchester, Manchester, UK}

\author{Alexander J. Healey} 
\thanks{These authors contributed equally to this work.}
\affiliation{School of Physics, University of Melbourne, Victoria 3010, Australia}
\affiliation{Centre for Quantum Computation and Communication Technology, School of Physics, University of Melbourne, Victoria 3010, Australia}
\affiliation{School of Science, RMIT University, Melbourne, Victoria 3001, Australia}

\author{Yi You} 
\affiliation{National Graphene Institute, The University of Manchester, Manchester, UK}
\affiliation{Department of Physics and Astronomy, The University of Manchester, Manchester, UK}

\author{Islay O. Robertson} 
\affiliation{School of Science, RMIT University, Melbourne, Victoria 3001, Australia}

\author{Sam C. Scholten} 
\affiliation{School of Physics, University of Melbourne, Victoria 3010, Australia}
\affiliation{Centre for Quantum Computation and Communication Technology, School of Physics, University of Melbourne, Victoria 3010, Australia}

\author{Ashok Keerthi} 
\affiliation{National Graphene Institute, The University of Manchester, Manchester, UK}
\affiliation{Department of Chemistry, The University of Manchester, Manchester, UK}

\author{Gary Harrison} 
\affiliation{Department of Materials, The University of Manchester, Manchester, UK}

\author{Lloyd C. L. Hollenberg} 
\affiliation{School of Physics, University of Melbourne, Victoria 3010, Australia}
\affiliation{Centre for Quantum Computation and Communication Technology, School of Physics, University of Melbourne, Victoria 3010, Australia}

\author{Boya Radha} 
\email{radha.boya@manchester.ac.uk}
\affiliation{National Graphene Institute, The University of Manchester, Manchester, UK}
\affiliation{Department of Physics and Astronomy, The University of Manchester, Manchester, UK}

\author{Jean-Philippe Tetienne}
\email{jean-philippe.tetienne@rmit.edu.au}
\affiliation{School of Science, RMIT University, Melbourne, Victoria 3001, Australia}


\begin{abstract} 

Magnetic van der Waals materials are often proposed for use in future spintronic devices, aiming to leverage the combination of long-range magnetic order and near-atomic thinness to produce energy-efficient components. One class of material that has been discussed in this context are the iron halides FeCl$_2$ and FeBr$_2$, which are A-type antiferromagnets with strong uniaxial magnetocrystalline anisotropy. However, despite characterization of the bulk materials, the possibility for sustaining the magnetic behaviors that would underpin such applications in thin flakes has not been investigated. In this work, we use nitrogen-vacancy (NV) center microscopy to quantitatively image magnetism in individual exfoliated flakes of these iron halides, revealing the absence of magnetic remanence, a weak induced magnetization under bias field and variable behavior versus temperature. We show that our results are consistent with the antiferromagnetic behavior of the bulk material with a soft ferromagnetic uncompensated layer, indicating that extended ($>1~\mu$m) ferromagnetic domains are not sustained even at low temperatures (down to 4~K). Finally, we find that the magnetic order is strongly affected by the sample preparation, with a surprising diamagnetic order observed in a thin, hydrated sample. 

\end{abstract}

\maketitle 


\section{Introduction}

\begin{figure*}
    \centering
    \includegraphics[width=0.8\linewidth]{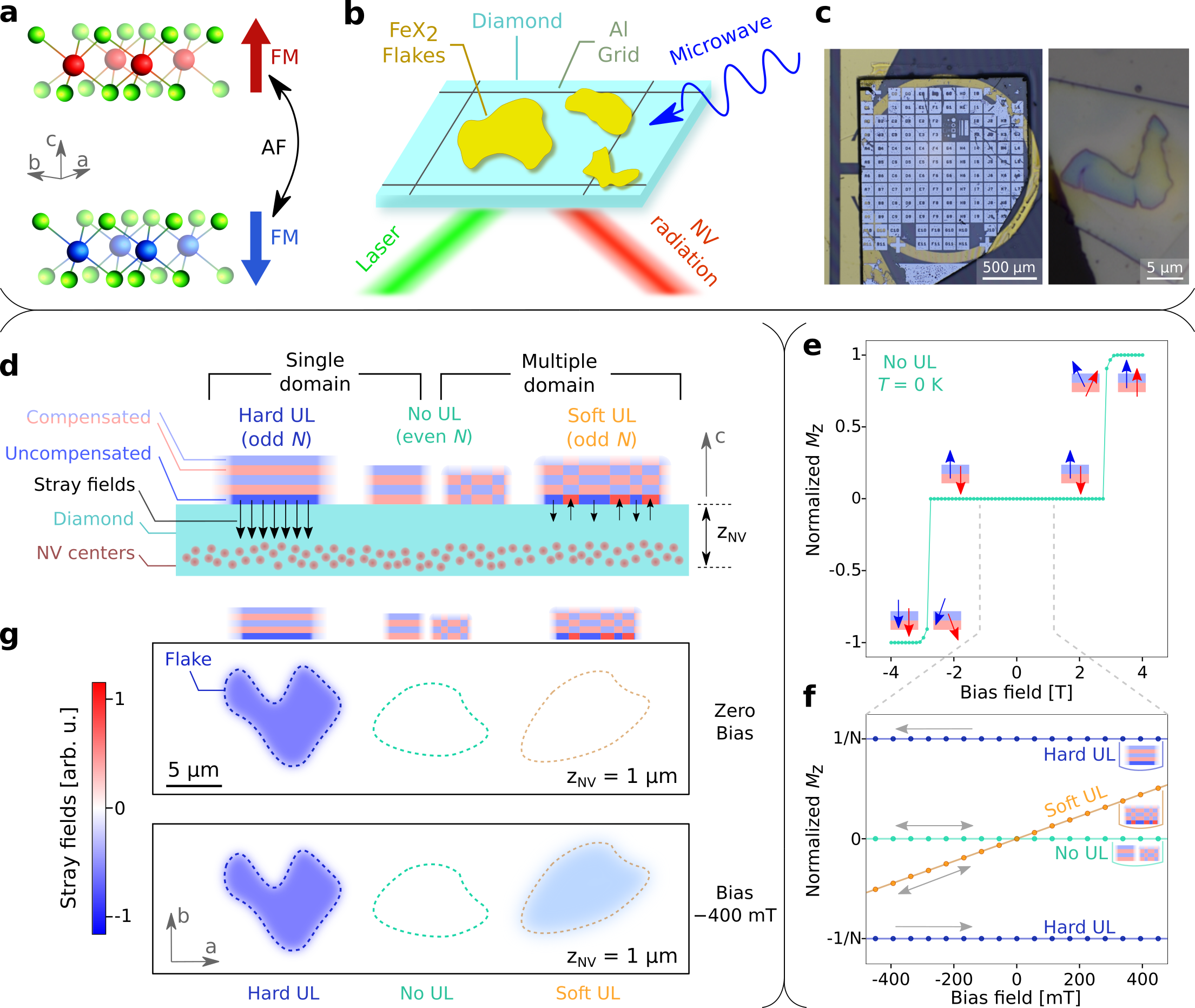}
    \caption{\textbf{Experimental setup and expectations for antiferromagnetic flakes}.
    \textbf{(a)} Crystallographic scheme for iron halides structures: the Fe$^{2+}$ ions (blue and red) are ferromagnetically coupled within a layer, and each layer is antiferromagnetically coupled with the adjacent ones. Magnetic contributions from the halogen ions (green) are negligible.
    \textbf{(b)} Widefield magnetometry experimental setup: the samples lie on top of a diamond substrate, which is patterned with an aluminum grid and an omega-shape gold resonator for microwave (MW) delivery. A green laser beam illuminates the diamond from below and the NV radiation is collected by a sCMOS camera.
    \textbf{(c)} Optical images of (left) the diamond substrate, where the MW resonator and aluminum grid can be seen, and (right) a FeBr$_2$ thin flake covered by a hBN flake.
    \textbf{(d)} Side view scheme of the thin flakes lying on top of the diamond, according to the different models from left to right: hard uncompensated layer (``Hard UL") for single domain configuration, ``No~UL" for both single and multiple domains and ``Soft UL" for multiple domains. The generated stray fields are sensed by a dense layer of NV spin defects at a standoff distance $z_{\rm NV}$ below the top surface.
    \textbf{(e)} Simulated hysteresis cycle at zero absolute temperature of the z-projected magnetization (normalized to the saturation value) as a function of the bias field applied along the $c$~axis, for an iron halide flake with $N$ layers according to the No~UL model in the single domain configuration. The schemes show the magnetization orientation of the sublattices at different stages.
    \textbf{(f)} Zoom-in version of the previous plot for our bias field measurement range,
    showing the models No~UL (green), Hard UL (blue) and Soft UL (orange). The arrows indicate the sweeping bias field direction.
    \textbf{(g)} Simulated stray fields for arbitrary flakes' geometries under zero bias (top panel) and $-400$~mT bias field (bottom panel). Each column corresponds to a different model from left to right: Hard UL, No~UL and Soft UL.
    }
    \label{fig1}
\end{figure*}

Intense research is directed towards magnetic van der Waals materials due to their potential for allowing the creation of miniaturised and energy-efficient spintronic devices~\cite{Sierra2021VanOpto-spintronics, Lin2019Two-dimensionalElectronics, Ashton2017}. One class of interesting materials in this context are van der Waals A-type antiferromagnets, which feature both ferromagnetic (FM) order within a single layer and antiferromagnetic (AF) coupling between adjacent layers [see \textcolor{blue}{Figure~\ref{fig1}(a)}], and thus resemble a naturally-occurring version of a synthetic antiferromagnet~\cite{Seo2021TunableMagnet} which finds widespread use in conventional spintronics~\cite{Duine2018SyntheticSpintronics}. The combination of these properties presents opportunities for exploiting the ferromagnetism of an extremal layer (for example to induce a proximity-based effect~\cite{Huang2020EmergentHeterostructures,Bora2021MagneticHeterostructure}) while minimising the total stray field emanating from the component, and hence limiting cross-talk or interference with other parts of a device~\cite{Jungwirth2016AntiferromagneticSpintronics}. 

The iron halides FeCl$_2$ and FeBr$_2$ are compounds known to exhibit A-type AF ordering in the bulk form~\cite{Jacobs1964,Jacobs1967MagneticFeCl2}, and have recently been synthesised in the ultrathin regime~\cite{Zhou2020AtomicallyEpitaxy,Cai2020FeCl2Effect}.
Further, it has previously been argued that monolayer FeCl$_2$ could possess a half-metallic ground state~\cite{Torun2015,Feng2020FeCl2/MoS2/FeCl2Applications} (although this is disputed~\cite{Yao2021FragileFeCl2}), while bilayer A-type antiferromagnets in general are known to become half-metallic when gated~\cite{Gong2018}. Crucial to many such spintronic applications (i.e.\ half-metallicity or interface effects) is the existence of extended FM domains within the monolayer near zero field and at accessible temperatures. 

These materials have previously been characterized in the bulk or powdered form using techniques such as neutron diffraction~\cite{Wilkinson1959NeutronCoCl2,Yelon1975,Binek2000}, which have established their AF behaviour, but have not addressed the thickness dependence of this order or important properties such as domain size and coercivity. Crucially, there is precedent for magnetic properties emerging in thin flakes of van der Waals materials that are not observed in the bulk material, such as hard ferromagnetism in Fe$_3$GeTe$_2$~\cite{Tan2018HardFe3GeTe2} and local perturbation of local A-type AF order~\cite{Thiel2019ProbingMicroscopy}. In this work we complement previous work using widefield nitrogen-vacancy center microscopy~\cite{Scholten2021WidefieldProspects}, which allows quantitative magnetic images of individual flakes to be collected rapidly and with sub-micrometer spatial resolution, potentially revealing a variety of magnetic phenomena, including FM and AF behaviours~\cite{Broadway2020ImagingFerromagnet,Healey2022VariedMicroscopy}. 

We study mechanically exfoliated, high purity flakes of varying thickness up to 100's of nm, and first find limited remanent magnetization at temperatures down to 4~K for both FeCl$_2$ and FeBr$_2$. We then observe diverse behaviour when varying the background magnetic field and sample temperature, sometimes even within the same flake. By comparing our results to micromagnetic models, considering the possibility for paramagnetic (PM) and AF ordering, we find that these observations are most consistent with soft ferromagnetism in uncompensated monolayers, unable to sustain large domains ($\gtrsim 1\,\mu$m) even at $T\approx4$\,K. In addition, we observed a surprising diamagnetic order in a thin, hydrated sample of FeBr$_2$ produced via a wet transfer method. Our results provide useful information on the magnetic behaviour of FeCl$_2$ and FeBr$_2$ in the thin, single flake regime, and highlight the importance of sample preparation on their magnetic order. 

\section{Experimental details}

The iron halides samples studied in this work were made from commercially sourced anhydrous bulk FeCl$_2$ (97\%, Acros organics) and FeBr$_2$ (98\%, Acros organics) materials, mechanically exfoliating thin flakes and transferring them onto a diamond substrate. We found that the as-bought iron halides are flaky materials that can be directly exfoliated from powders using a tape, without any solid state crystal growth, obtaining large enough flakes (few tens of microns in size) down to tens or few hundreds nanometers thin. All exfoliation and transferring steps of flakes were carried out in a glovebox at room temperature by the dry transfer method (unless specified otherwise). The flakes were mechanically exfoliated onto polymethylsiloxane (PDMS) sheets and subsequently transferred onto a specified region of a diamond substrate. In order to protect the samples against ambient degradation, the iron halides flakes were sealed with a hBN flake (thickness $<40$~nm). More details about the complete process can be found in \textcolor{teal}{Appendix~\ref{apx_stru_char}}.

The experimental setup for widefield magnetometry is depicted in \textcolor{blue}{Figure~\ref{fig1}(b,c)}: the iron halides flakes lie on top of a diamond substrate, which is patterned with an aluminum grid to protect the samples from laser illumination, and an omega-shaped gold resonator for microwave (MW) delivery. A green laser beam ($\lambda=532$~nm) illuminates the diamond from below, exciting a dense layer of nitrogen-vacancy (NV) spin defects $\approx500$~nm below the top surface (standoff distance). These defects are sensitive to the environmental magnetic fields and radiate photoluminescence ($\lambda\approx650$~nm) that is collected by a sCMOS camera~\cite{Scholten2021WidefieldProspects}. The stray magnetic fields, which relate to the sample magnetization, can be mapped in a 50~$\times$~50~$\mu$m$^2$ field of view, with $\approx$~1~$\mu$m resolution. All the experiments were performed in a dry cryostat (Attocube), with working temperatures down to 4~K and equipped with a superconducting vector magnet which can supply a bias magnetic field up to 1~T in any direction. See \textcolor{teal}{Appendix~\ref{apx_NV_widefield}} for a more explicit description.


\section{Theoretical expectations}

Widefield NV microscopy is a stray field measurement technique, meaning that we aim to infer information about a sample's magnetization by measuring the resulting magnetic field~\cite{Doherty2013TheDiamond, Steinert2010HighDiamond}. The field measured is the projection along the NV axis, which for the following measurements is aligned with the anisotropy axis (the $c$~axis) of the iron halides samples. Before proceeding further, we begin with an overview of the possible limiting magnetic behaviours the iron halides may exhibit, and introduce how these will be distinguishable using NV microscopy. 

Both FeBr$_2$ and FeCl$_2$ materials are expected to have A-type AF order below the N\'eel temperature $T_{\rm N}$ (14~K for FeBr$_2$, 24~K for FeCl$_2$) and under low bias fields ($<$1~T) applied along the $c$~axis, alternating positive and negative magnetization for consecutive layers~\cite{Jacobs1964, Jacobs1967MagneticFeCl2, Stryjewski1977, Pasternak1976, Pelloth1995, McGuire2017, Yang2021}, as schemed in \textcolor{blue}{Figure~\ref{fig1}(a)}. For bulk crystals, it is expected that each layer behaves as a single FM domain, with all magnetic moments pointing in the same direction, then a pair of adjacent layers is magnetically compensated and generates no stray fields, while an uncompensated layer (UL) contributes with its single domain magnetization. We call these single domain models ``No~UL" when the number of layers~$N$ is even and hence the magnetization is completely compensated, and ``Hard UL" when $N$ is odd and there is a single domain UL. On the other hand, in thin and small flakes comprising not more than a few hundreds of layers and tens of $\mu$m for the in-plane dimensions, the single domain assumption could be relaxed because crystal imperfections, finite temperature and shape factors will compete with the global AF interactions. As a result, adjacent layers will still compensate each other in pairs, but an UL may exhibit a multi-domain structure, resembling a soft ferromagnet. We call this model ``Soft UL" when there is an UL (odd~$N$); while the even~$N$ case has the same stray field predictions as the ``No~UL" model, independently of the domain configuration, and so we will consider them equivalent. Notice that the generated stray fields are very different according to each model, but they are all described by a general AF behavior; the difference lies in the presence or not of an UL and, if any, its single or multi-domain structure.

The stray magnetic fields generated by arbitrary flakes' geometries can be calculated if the material's magnetization~$\vec{M}$ is known, as described elsewhere~\cite{Broadway2020ImprovedMeasurements}. In a single domain model, we can consider each FeBr$_2$ or FeCl$_2$ layer within the flake as a unit in which all magnetic moments~$\vec{\mu}$ are pointing in the same direction (then saturated). Using literature values for $\mu$~\cite{Wilkinson1959NeutronCoCl2} and the lattice parameters~\cite{Balucani1985}, we obtained the saturation magnetization $M_{\rm S}$ for FeBr$_2$ (0.49~MA/m) and FeCl$_2$ (0.61~MA/m). Having determined $|\vec{M}|=M_{\rm S}$ for each layer, the magnetization orientations can be simulated by considering them as free variables in an energy density~$E$ equation that has to be minimized:
\begin{equation}
    E = E_{\rm Zeeman} + E_{\rm demag} + E_{\rm magcrys} + E_{\rm AF} \, ,
\label{eq_E}    
\end{equation}

\noindent where the energy density contributions come from the Zeeman, demagnetizing, magnetocrystalline and antiferromagnetic exchange terms, respectively. For A-type antiferromagnetism, a flake composed by $N$ layers can be treated as two sublattices (namely ``odd" and ``even") with independent magnetization orientations. We consider bias fields $B_0$ applied along the $c$~axis (also $z$ direction) and define the polar angles $\theta$ with respect to this axis. For the general case of axial symmetry, the only two free variables to minimize $E$ are the polar angles $\theta_{\rm odd}$ and $\theta_{\rm even}$. Each energy contribution can be described as follows:
$$ E_{\rm Zeeman} = -B_0 M_{\rm S} \, \left[\frac{N_{\rm odd} \cos(\theta_{\rm odd}) + N_{\rm even} \cos(\theta_{\rm even})}{N}\right] \, , $$
$$ E_{\rm demag} = \frac{1}{2}\mu_0 M_{\rm S}^2  \, \left[\frac{N_{\rm odd} \cos(\theta_{\rm odd}) + N_{\rm even} \cos(\theta_{\rm even})}{N}\right]^2 \, , $$
$$ E_{\rm magcrys} = K_{\rm u} \, \left[\frac{N_{\rm odd} \sin^2(\theta_{\rm odd}) + N_{\rm even} \sin^2(\theta_{\rm even})}{N}\right] \, , $$
$$ E_{\rm AF} = J_{\rm AF} \left[\frac{N-1}{N} \cos(\theta_{\rm odd}-\theta_{\rm even})\right] \, , $$

\noindent where $M_{\rm S}$ is the magnetization saturation, $N_{\rm odd}$ and $N_{\rm even}$ are the number of layers for each sub-lattice, $K_{\rm u}$ is the uniaxial anisotropy and $J_{\rm AF}$ is the AF exchange constant (absolute value). For more details about the explicit formulation and parameter values see \textcolor{teal}{Appendix~\ref{apx_Edens}}.

The simulated hysteresis cycle for a FeBr$_2$ flake according to the No~UL model (even~$N$) at temperature $T=0$~K is shown in \textcolor{blue}{Figure~\ref{fig1}(e)}, where the normalized $z$-projection of the total flake magnetization~$M_{\rm z}/M_{\rm S}$ is a function of the bias field $B_0$. The schemes show the magnetization orientation of the sublattices at different stages. For $|B_0| \gtrsim 3$~T, the Zeeman energy overcomes the AF exchange energy and the sample reaches magnetic saturation following a spin-flip transition~\cite{Pelloth1995}. For lower biases, the AF exchange prevails and the two sublattices are magnetically compensated, meaning $M_{\rm z} = 0$. The Hard UL model is essentially similar, although the magnetization is not fully compensated in the low bias region because of the UL, which contributes with a fraction $1/N$ of the normalized magnetization.

The cycle is zoomed-in in \textcolor{blue}{Figure~\ref{fig1}(f)} for our typical measurement range, including the three models: Hard UL (blue), No~UL (green)  and Soft UL (orange), which we will explain later. The gray arrows show the bias field sweeping direction, which evidences reversible processes for the No~UL and Soft UL models, whereas the magnetization in the Hard UL model depends on the previous exposure to the bias field. Examples for the computed stray fields produced by arbitrary flakes' geometries and sensed by the NV centers in a parallel plane at $z_{\rm NV}=1 \, \mu$m, as illustrated in \textcolor{blue}{Figure~\ref{fig1}(d)}, are shown in \textcolor{blue}{Figure~\ref{fig1}(g)}, for zero bias (upper panel) and $B_0=-400$~mT (lower panel). As expected for the Hard UL model, a single domain UL generates a constant stray field in this low bias regime (left column), where for this particular case we proposed a previous saturated magnetization in the $-z$ orientation. For the No~UL model (central column), the magnetization is totally compensated, hence no stray fields are generated. We note that these simulations are valid in the $T=0$~K limit, and we expect a weak linear response to the bias field due to the thermal energy at finite temperatures.

The multiple domain models (No~UL and Soft UL) assume that the domain size is below our 1~$\mu$m spatial resolution, in which case the detected stray fields would be averaged over the contributions of many domains within the same flake. In order to simulate the energy landscape according to this model, \textcolor{blue}{Equation~\ref{eq_E}} should be updated by adding complex domain dynamics, which depend on many factors such as the sample's geometry and crystallographic defects. In contrast with the single domain models, we expect a strong dependence on the bias field, since magnetic domains will be competing to grow and at the same time to balance the AF exchange. In the case of multi-domain compensated layers (No~UL), no stray fields will be generated regardless of the domain dynamics. However, simulating the UL case (Soft UL) under these conditions is a nontrivial task. To circumvent this complicated scenario we consider a soft FM model [see cartoon in \textcolor{blue}{Figure~\ref{fig1}(d)}], which depends linearly on the bias field and has no measurable remanent magnetization, essentially resembling a monolayer soft ferromagnet. For low bias fields, the hysteresis loop is linear and reversible, as shown in \textcolor{blue}{Figure~\ref{fig1}(f)} (orange curve). The corresponding stray fields are simulated in \textcolor{blue}{Figure~\ref{fig1}(g)} (right column), showing no remanent magnetization for zero bias (top panel) and a non-saturated value for low bias (bottom panel).

  
\section{Results}

\begin{figure*}
    \centering
    \includegraphics[width=0.8\linewidth]{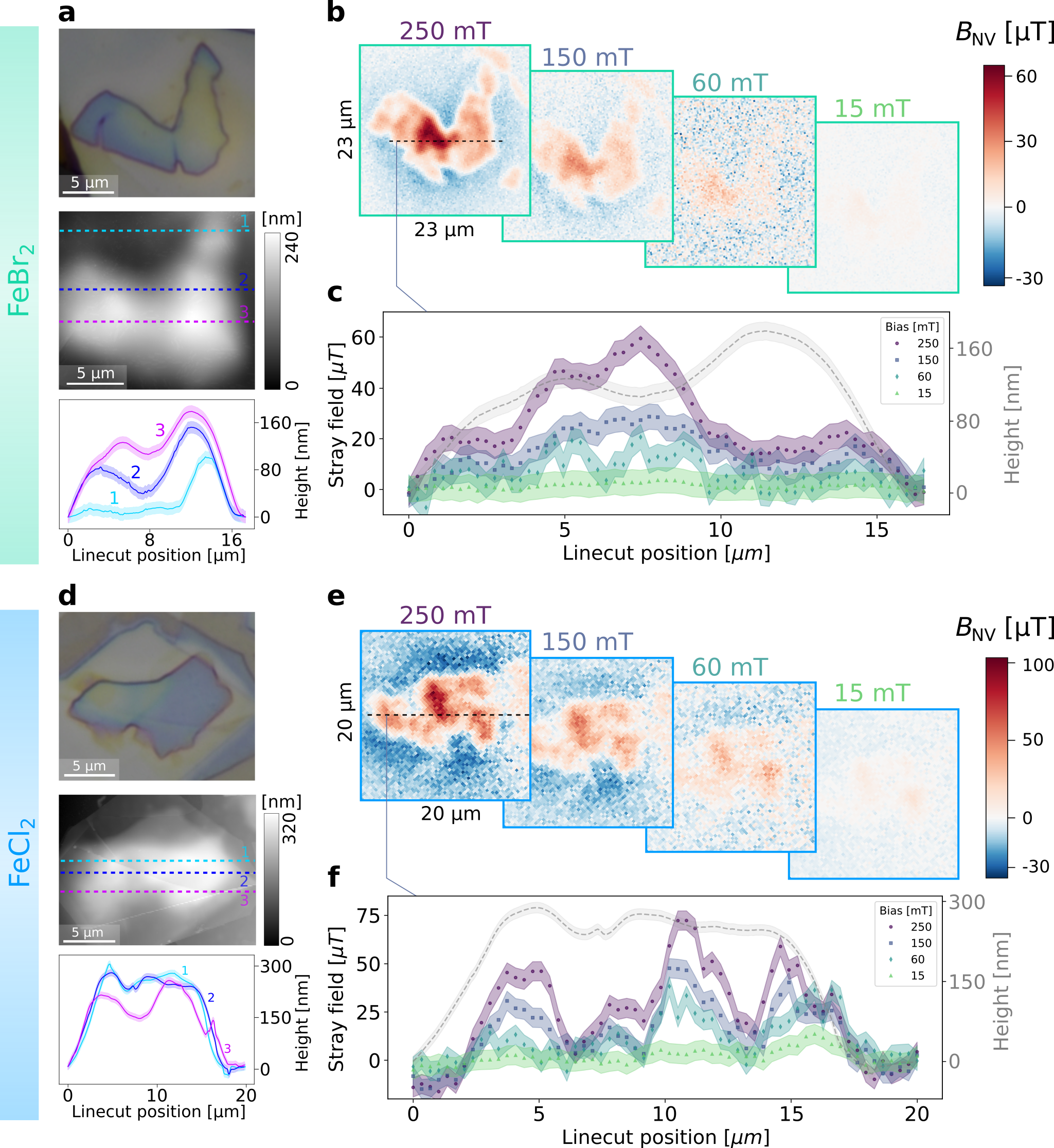}
    \caption{\textbf{Samples characterization and magnetic maps showing negligible remanent magnetization}.
    \textbf{(a)},\textbf{(d)} Optical and AFM characterization for FeBr$_2$ and FeCl$_2$ samples, respectively. Optical images (top panels) match the AFM measurements (middle panels), in which linecuts are correlated with height profiles (bottom panels, shaded areas represent experimental uncertainties).
    \textbf{(b)},\textbf{(e)}: Stray field map series for FeBr$_2$ and FeCl$_2$ samples, respectively, under different bias fields. For both samples, the magnetic signal decreases with the bias field, pointing to zero or very low remanent magnetization. Linecuts on the left-most maps (matching one of the AFM linecuts) are correlated with the \textbf{(c)},\textbf{(f)} stray fields profiles for FeBr$_2$ and  FeCl$_2$ samples, respectively, evidencing no correlation with the samples' topographies.
    }
    \label{fig2}
\end{figure*}

Optical images of the iron halides samples are shown in \textcolor{blue}{Figure~\ref{fig2}(a,d)} for both FeBr$_2$ and FeCl$_2$ flakes, respectively, covered by protective hBN flakes. Atomic force microscopy (AFM) images are in the middle panels, with linecuts linked to the height profiles in the bottom panels. For both samples, the flakes are not uniform in height and are composed of many layers, ranging from 100~to 200~nm (160~to 320 layers) for FeBr$_2$ and 150~to 300~nm (260~to 510 layers) for FeCl$_2$ (calculations are based on the lattice parameters~\cite{Balucani1985}).

We performed a set of widefield magnetometry measurements at different bias fields applied along the $c$~axis, keeping the temperature~$T$ constant at 4~K. Another set was measured with a constant 200~mT bias field and varying the temperature up to 50~K. The first set ($T=4$~K) is displayed as stray field maps series in \textcolor{blue}{Figure~\ref{fig2}(b,e)} for FeBr$_2$ and FeCl$_2$ samples, respectively. We observe that in both samples the stray magnetic fields vary reversibly with the bias field, showing no signs of remanent magnetization or hysteresis (we show this explicitly in \textcolor{teal}{Appendix~\ref{apx_NoHyst}}). This approximately linear response is in principle compatible with the Soft UL model, with almost zero coercivity, or PM behavior. By comparing linecuts across the magnetic and AFM images [\textcolor{blue}{Fig.~\ref{fig2}(c,f)} and \textcolor{blue}{Fig.~\ref{fig2}(a,d)}] for the different bias fields, we can see that the magnetic signals are not strongly correlated with the height profiles, as would be expected for PM. This observation is the first hint that the Soft UL model may best describe the systems, which supposes that domains below the measurement spatial resolution reversibly grow in size with applied magnetic field strength. However, establishing this quantitatively requires a more thorough comparison to theory as we will introduce in the next section.


\subsection{Analysis of magnetic field dependence}

Following the existing characterization of the bulk material~\cite{Jacobs1964, Jacobs1967MagneticFeCl2, Di2004, Ropka2001, Binek2000, Pelloth1995, Alben1969, Pasternak1976, Yang2021, Stryjewski1977, Baltz2018}, we would expect our measurements at $T=4$~K ($T< T_{\rm N}$ for both materials) to show AF behavior in better agreement with our single-domain models. This would imply that the stray fields should remain constant either close to zero (No~UL) or at a fixed value (Hard~UL), but our experimental results contradict this interpretation. Conversely, the multiple domain Soft UL model may explain the non-uniform and not-constant behavior, evidencing an irregular distribution of magnetic domains, which will be mostly related to crystallographic defects and the sample geometry. However, a weak response with zero remanent magnetization could also be attributed to paramagnetism.

To consider the paramagnetic explanation, we build a model for thin flakes assuming that they are PM throughout their volume, in which case the magnetic response will be proportional to the number of layers. Each Fe$^{2+}$ ion (Br or Cl contributions are negligible) will contribute to the magnetization~$M_{\rm PM}$ based on their quantum numbers $S$~(spin), $L$~(orbital angular momentum) and $J$~(total angular momentum), see \textcolor{teal}{Appendix~\ref{apx_PM_model}} for more details. High quantum numbers will result in large~$M_{\rm PM}$, and for this model we use the lowest configuration $J=S=1$, $L=0$, as suggested in~\cite{Pelloth1995} (later we will demonstrate that even the lowest PM expectation is too high compared to the experimental values). The magnetization of the whole flake at temperature~$T$ and under a bias field~$B_0$ can be calculated as~\cite{Kittel1957IntroductionPhysics}: 
\begin{equation}
    M_{\rm PM} = n g J \mu_{\rm B} B_{\rm J}(x)\, ,
    \label{eq_PM}    
\end{equation}
$$ x = g J \mu_{\rm B} B_{0} / k_{\rm B} T \, ,$$

\noindent where $n$ is number of atoms per unit volume, $g=2$ is the Land\'e factor for this quantum configuration, $\mu_{\rm B}$ is the Bohr magneton, $B_{\rm J}$ is the Brillouin function and $k_{\rm B}$ is the Boltzmann constant. The magnetization $\vec{M}_{\rm PM}$ will be oriented parallel to the bias field direction, and the stray fields outside the flake can be simulated~\cite{Broadway2020ImprovedMeasurements}.

\begin{figure*}
    \centering
    \includegraphics[width=0.8\textwidth]{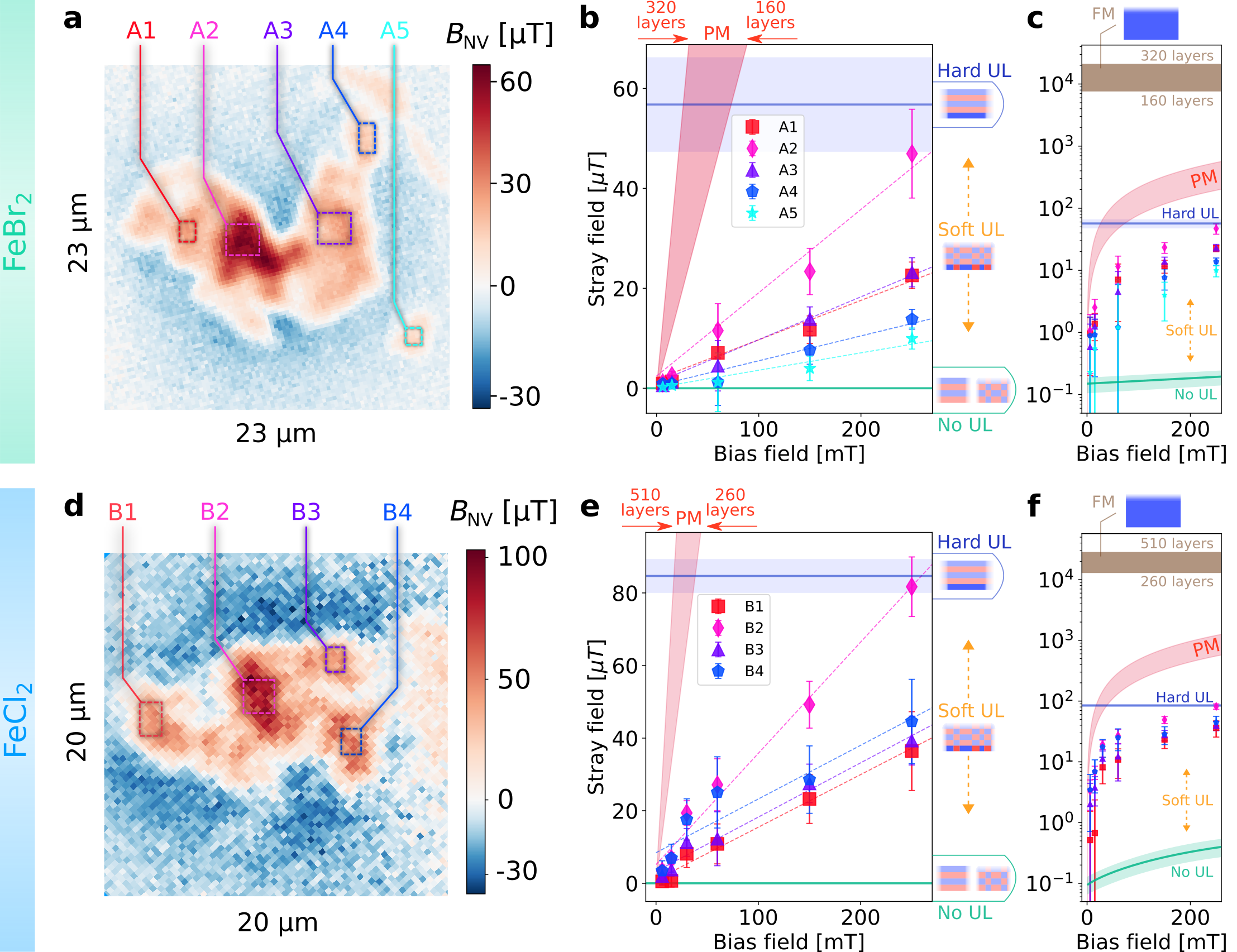}
    \caption{\textbf{Magnetic signal versus bias field along with possible magnetic models}.
    \textbf{(a,d)} Stray field maps at temperature 4~K and bias 200~mT along the $c$~axis. The color scale represents the field intensity sensed by the NV sensors at 500~nm (FeBr$_2$) and 250~nm (FeCl$_2$) standoff distance.
    \textbf{(b,e)} Averaged stray fields as a function of bias field for identified regions (error bars represent one standard deviation) along with expectations for the different models: No~Uncompensated Layer (green solid line), Hard Uncompensated Layer (blue solid line, the shaded area represents one standard deviation), Soft Uncompensated Layer, which represents a region in-between as depicted by the right-margin schemes, and Paramagnetic (red shaded area). Dashed curves are linear fits for experimental points.
    \textbf{(c,f)} Same plots but using logarithmic scale for stray fields and including the ferromagnetic model (brown shaded area).
    }
    \label{fig3}
\end{figure*}

Considering the Soft UL and PM models, we analyze the magnetic properties in small regions within the thin flakes, looking for evidence that supports any of them. \textcolor{blue}{Figure~\ref{fig3}(a,d)} identifies several regions for FeBr$_2$ and FeCl$_2$ samples, respectively, in which the stray fields are spatially averaged. The results are shown in \textcolor{blue}{Figure~\ref{fig3}(b,e)} (error bars represent one standard deviation in the spatial average), along with the No~UL model (green line at 0~$\mu$T, which is extrapolated from bulk data -- where the effect of any UL is negligible -- taken at 4~K~\cite{Jacobs1964,Jacobs1967MagneticFeCl2}), the Hard UL model (blue constant line at the top, the shaded area represents one standard deviation) and the PM model for a wide range of number of layers according to the AFM measurements (red shaded area). The Soft UL model is in between the No~UL and Hard UL ones, as shown by the schemes on the right margin, the precise scaling is difficult to model but will involve a monotonic rise towards saturation within the layer, featuring negligible remanence and coercivity.

In both samples and in all regions, the stray fields exhibit a linear dependence on the bias field intensity (dashed fitted lines), with negligible remanent magnetization and coercivity. As discussed before, this general behavior is in clear disagreement with the single domain models in this low bias field regime, showing either zero stray fields (No~UL) or constant values for an UL (Hard UL) , equal to (57$\pm$9)~$\mu$T for FeBr$_2$ and (85$\pm$4)~$\mu$T for FeCl$_2$ (based on anticipated saturation magnetizations 0.49~MA/m and 0.61~MA/m, respectively, see \textcolor{teal}{Appendix~\ref{apx_Edens}}). Even if we consider a mixed case for our spatial average, containing sub-micrometer regions with even and odd number of layers, the average stray field would have an intermediate value but it should be constant as a function of the bias field. 

A pure PM model taking into account lower and upper boundaries for the number of layers (obtained from the AFM measurements) also fails, because the experimental magnetic signal should be much more intense. \textcolor{blue}{Figure~\ref{fig3}(c,f)} reproduces the previous plots but using logarithmic scale for the stray fields to show that the expected PM response is roughly ten times stronger. To have the complete picture, a full FM model (brown shaded area) is included taking the same layer boundaries as before, showing that aligning all the magnetic moments in the flake would generate a magnetic response more than one order of magnitude higher compared to the PM model at this bias field range. 

Despite the fact that pure PM is not the appropriate model to describe our samples, it is possible that only small volumes within a flake are PM, contributing with a weak linear response. However, the lack of correlation between the flakes' topography [i.e.\ the height distribution; \textcolor{blue}{Figure~\ref{fig2}(a,d)}] and the stray fields intensity suggests that multiple (sub-diffraction limit) domain structures are responsible for the inhomogeneous behavior, at least in addition to any of these small PM volumes. The Soft UL model fits this description and predicts a linear and weak response resulting from the expansion of domains, in principle a fraction of the stray fields generated by a single UL that increases as the bias field is increased. Nevertheless, because we expect many crystallographic defects and stacking faults throughout a flake, it may be more likely that a sample has a few effectively uncompensated layers rather than a single one. Given that there is no strong correspondence between the strength of the magnetic signal and the sample thicknesses measured by AFM, a more likely explanation is that the interlayer antiferromagnetism can break down locally as has been observed in similar A-AF systems~\cite{Thiel2019ProbingMicroscopy,Healey2022VariedMicroscopy}. The interlayer antiferromagnetism could potentially be disrupted by variable stacking orders, stacking faults, or even crystal strain.

 
\subsection{Temperature dependence}

The Soft UL and PM hypotheses can be further examined by looking at the temperature-dependent volume magnetic susceptibility $\chi_{\rm v}(T)$. Since we previously showed that the stray fields are proportional to the bias fields $B_0$ at $T=4$~K in our measurement range, we will assume the following approximation:
$$ \chi_{\rm v} = \frac{\partial (\mu_0 M)}{\partial B_0} \approx \frac{\mu_0 M}{B_0}\, , $$

\noindent where $\mu_0$ is the vacuum permeability and $M$ is the sample's magnetization. In the PM case, $\chi_{\rm v}^{\rm PM}(T)$ can be calculated using $M=M_{\rm PM}(T)$ from \textcolor{blue}{Equation~\ref{eq_PM}}. Modelling or simulating the Soft UL susceptibility~$\chi_{\rm v}^{\rm sUL}(T)$ would require information about the magnetic domains, which is beyond the scope of this work. However, we can guess that a lower bound would be the single domain (SD) AF susceptibility~$\chi_{\rm v}^{\rm SD}(T)$ (see \textcolor{teal}{Appendix~\ref{apx_AF_MvsT_expec}}), because alternating magnetic layers would try to hold their AF structure regardless of the bias field, whereas multiple domains in an UL would be more responsive to the bias field. Following this argument, we expect that the sensitivity to the bias field is mainly dictated by a single UL; while a full FM susceptibility~$\chi_{\rm v}^{\rm FM}(T)$, in which all the layers are contributing, would be a much higher bound.

\begin{figure*}
    \centering
    \includegraphics[width=0.8\textwidth]{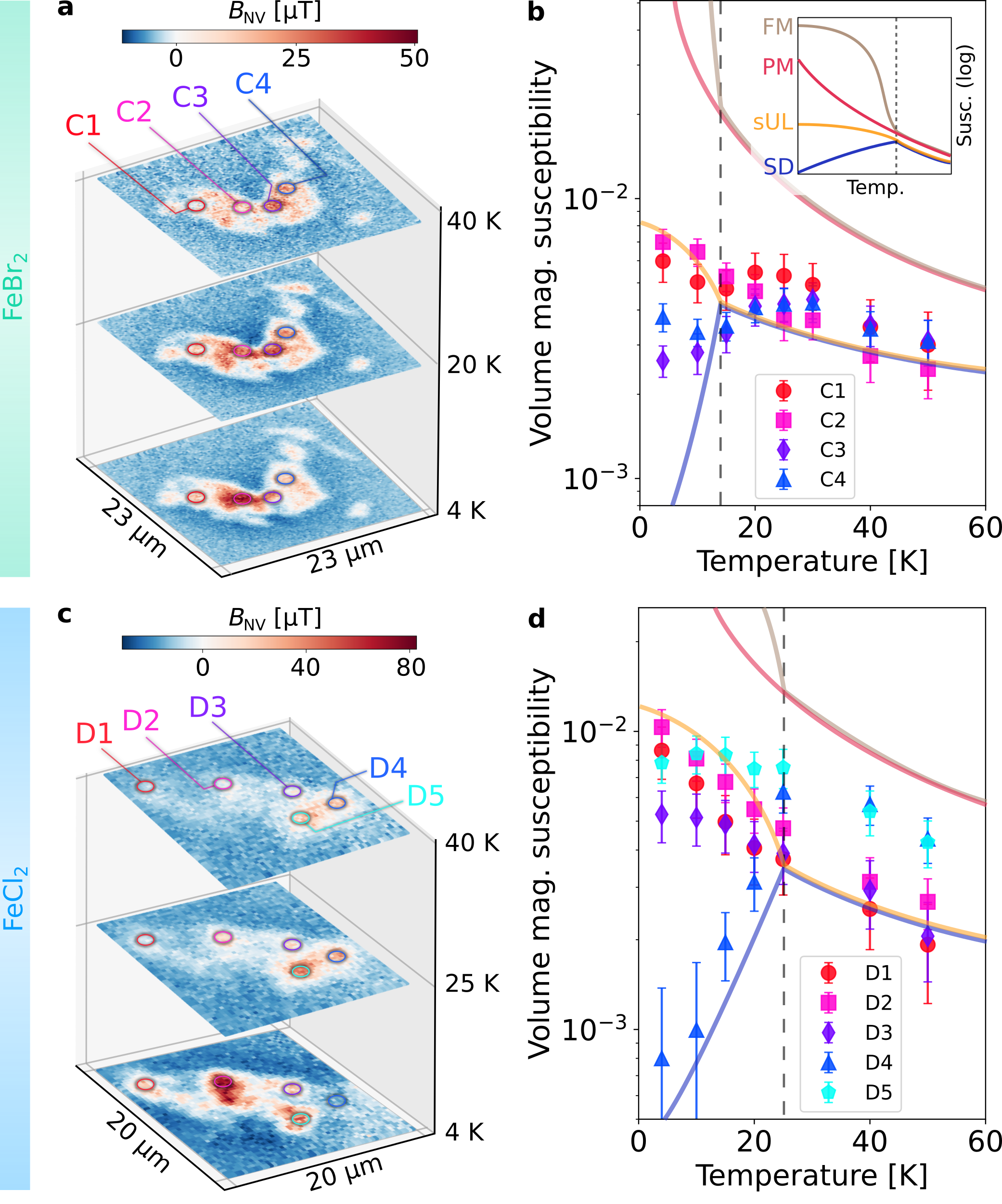}
    \caption{\textbf{Volume susceptibility versus temperature along with models expectations}.
    \textbf{(a,c)} Stray field maps under 200~mT bias along the $c$~axis, at temperatures below, close and above the N\'eel temperature. The color scale represents the field intensity sensed by the NV sensors at 500~nm (FeBr$_2$) and 250~nm (FeCl$_2$) standoff distance.
    \textbf{(b,d)} Averaged volume magnetic susceptibility as a function of temperature for identified regions (error bars represent one standard deviation) along with expectations according to different models (solid lines): 
    Single Domain antiferromagnetism (blue), Soft Uncompensated Layer (orange), Paramagnetism (red) and Ferromagnetism (brown). The inset in~\textbf{(b)} shows the qualitative behaviors of these models in a larger scale, valid for both samples. The N\'eel temperature is represented by a vertical dotted line.
    }
    \label{fig4}
\end{figure*}

The experimental stray fields maps are displayed in \textcolor{blue}{Figure~\ref{fig4}(a,c)} for FeBr$_2$ and FeCl$_2$ samples, respectively, for different temperatures and under a constant bias field $B_0 = 200$~mT along the $c$~axis. At first glance, it becomes clear that the flakes do not evolve as a whole unit: the magnetic signal decreases or increases with temperature depending on the region. We studied $\chi_{\rm v}(T)$ for representative regions, as shown in \textcolor{blue}{Figure~\ref{fig4}(b,d)} along with the N\'eel temperature (vertical dashed lines) and expectations for the different models. Error bars for experimental values were assigned by propagating the uncertainties from the spatial average of stray fields and volume estimation. Additionally, we include a qualitative plot for susceptibility expectations on a larger scale, see inset in \textcolor{blue}{Figure~\ref{fig4}(b)}, which is valid for both samples.

The magnetic models are mainly different below $T_{\rm N}$, where the interactions between magnetic moments are stronger than the thermal energy. Above $T_{\rm N}$, $\chi_{\rm v}(T)$ decays roughly as $1/T$ for all cases, slightly modified for the AF models (SD and Soft~UL) by the Weiss temperature $\theta_{\rm W}$ which gives $\chi_{\rm v}(T) \propto 1/(T+\theta_{\rm W})$ (see \textcolor{teal}{Appendix~\ref{apx_AF_MvsT_expec}} for details). Comparing the experimental results with the model predictions at low temperatures in \textcolor{blue}{Figure~\ref{fig4}(b,d)}, we observe that the PM expectations are at least one order of magnitude higher. We can conclude that although it is possible to have small PM volumes within the flake, paramagnetism is definitely not the global magnetic ordering.

The AF models describe the susceptibility much better, as expected for these materials, anticipating an increasing $\chi_{\rm v}(T)$ as $T$ decreases for a multiple domain structure and a decreasing $\chi_{\rm v}(T)$ in the single domain case. If we assume a multiple domain structure and the presence of uncompensated layers in these thin flakes, then it is reasonable to find average susceptibility values in between the SD and Soft~UL predictions, depending on each region. For instance, region~$D4$ [see \textcolor{blue}{Figure~\ref{fig4}(d)}] suggests the presence of a relatively large single domain while the other regions are compatible with sub-micrometer multiple domains.

The temperature results reinforce the previous discussion about the linear response to the bias field, in favor of the Soft UL model. Furthermore, the region-dependent behavior is in agreement with the hypothesis that crystallographic defects and irregular shape effects will play a major role and generate a multiple domain magnetic structure.


\subsection{Diamagnetic order in hydrated samples}

Finally, we examine a separate FeBr$_2$ sample fabricated using a wet transfer method, anticipating that the sample preparation may affect the magnetic behavior. Additionally, this sample was prepared on a $\langle$100$\rangle$ surface-oriented diamond sensor, meaning that all the NV axes have both parallel and perpendicular components relative to the sample $c$~axis, see \textcolor{teal}{Appendix~\ref{apx_dia_NVprojections}}. In this geometry, bias fields can be applied along any of the NV axes, which are not aligned with the anticipated crystal anisotropy axis, possibly allowing us to uncover more varied magnetic behaviors such as canting away from the $c$~axis~\cite{Healey2022VariedMicroscopy}. In particular, we are able to assess the degree to which the sample magnetization is pinned to the $c$~axis. 

\begin{figure}
    \centering
    \includegraphics[width=0.45\textwidth]{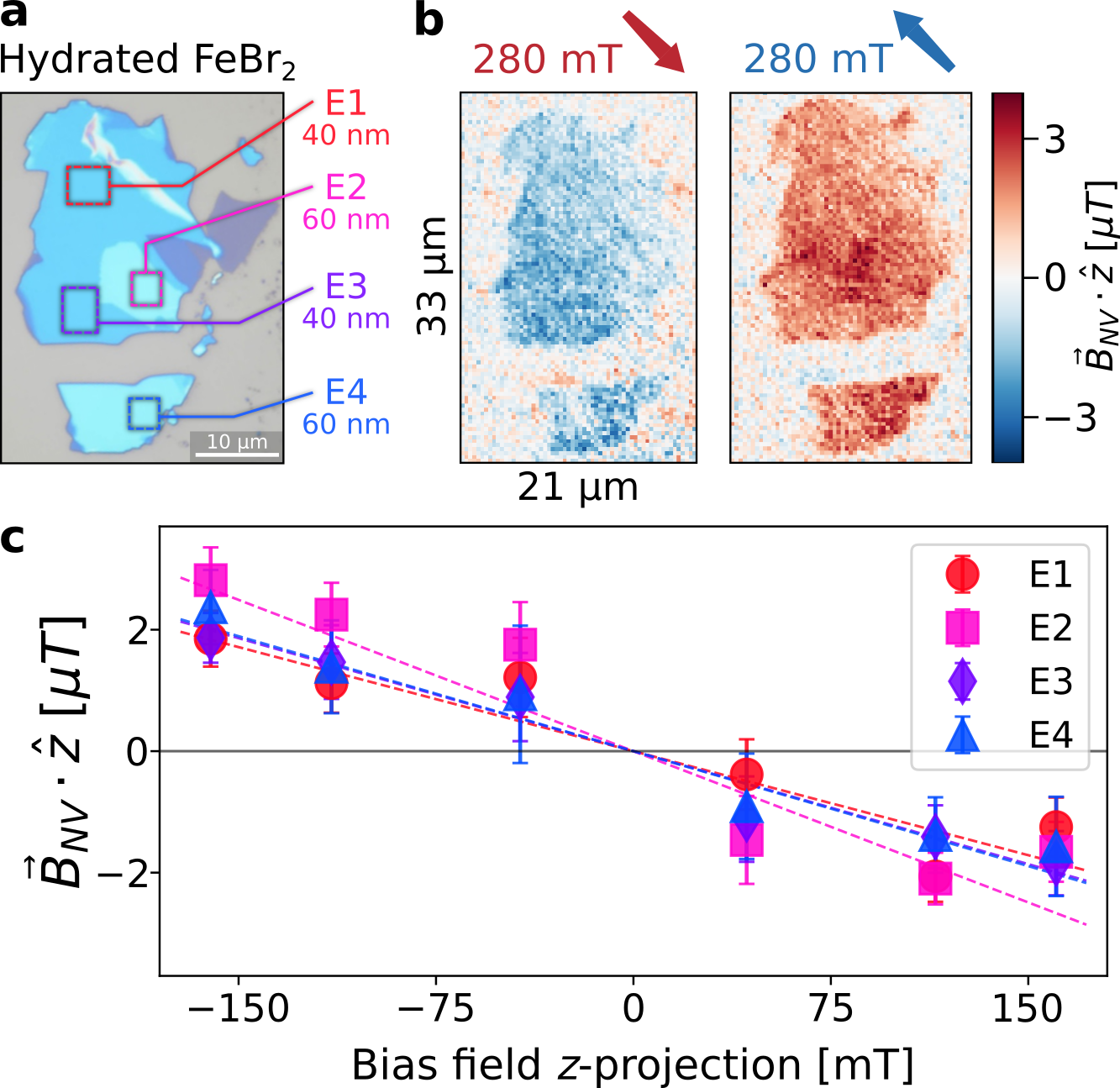}
    \caption{\textbf{Diamagnetic behavior in hydrated FeBr$_2$ sample}.
    \textbf{(a)}~Optical image in which several regions are identified and labeled according to their height.
    \textbf{(b)}~Reconstructed stray field maps measured at 4~K and projected along the $z$~axis for opposite 280~mT (left) and $-$280~mT (right) bias fields applied along the (-1,1,-1) NV axis.
    \textbf{(c)}~Average stray fields vs bias fields, both projected along the $z$~axis, for all identified regions (error bars represent one standard deviation uncertainty). A proportional and opposite response relative to the bias field can be observed in all regions, consistent with a diamagnetic ordering.
    }
    \label{fig5}
\end{figure}

\textcolor{blue}{Figure~\ref{fig5}(a)} shows an optical image of the sample, in which several regions are identified and labeled with their average height. The reconstructed stray field maps measured at 4~K and projected along the $z$~axis (for ease of comparison with our previous measurements) are shown in \textcolor{blue}{Figure~\ref{fig5}(b)} for opposite 280~mT and $-$280~mT bias fields applied along the (-1,1,-1) NV axis. Compared to our previous results we observe only very weak magnetic signals, at least one order of magnitude smaller. More surprisingly still, we observe that the detected average stray fields are opposite to the bias field, suggesting a diamagnetic order.

The stray fields in the selected regions are averaged, projected into the $z$~direction ($\vec{B}_{\rm NV} \cdot \hat{z}$) and plotted as a function of the bias field $z$-component, see \textcolor{blue}{Figure~\ref{fig5}(c)}. We observe a linear and negative response with similar values for all regions, independently of their height. Using linear fittings and correlating the stray fields with magnetization values, we find diamagnetic volume susceptibilities ranging from $-1\times 10^{-6}$ to $-2\times 10^{-6}$, consistent within the order of magnitude expected for a diamagnetic material. Through analysis of the stray field patterning and comparison with simulation (see \textcolor{teal}{Appendix~\ref{apx_dia_sims}}), we establish that strong anisotropy along the $c$~axis is still present, indicating that the magnetocrystalline anisotropy remains even while the usual magnetic order has disappeared.

\section{Conclusions}

Using widefield NV microscopy, we have measured the magnetic properties of individual FeCl$_2$ and FeBr$_2$ thin flakes (100-300~nm thick) over a range of bias and temperature conditions, complementing the previous bulk experimental research for these materials with novel results. No flakes examined exhibited magnetic remanence or obvious hysteresis, which we explain by a multiple domain model with domain sizes below our $\approx 1~\mu$m resolution and the presence of uncompensated layers. The volume magnetic susceptibility as a function of temperature showed good agreement with our model and suggested irregular domain configurations, which could originate from crystallographic defects and shape effects in these thin flakes. More surprisingly, we found that one sample was diamagnetic, which we tentatively attribute to hydration of the crystal during a wet transfer process. Our results show that high purity, iron halides thin flakes produced via a dry transfer method exhibit antiferromagnetic properties aligned with those observed in the bulk materials, but they do not sustain the large magnetic domains required for many spintronic applications. Future investigations into methods better suited to producing high quality monolayer and bilayer samples such as molecular beam epitaxy~\cite{Zhou2020AtomicallyEpitaxy,Cai2020FeCl2Effect} may be worthwhile to establish whether larger domains can exist in the ultrathin regime.


\begin{acknowledgments}

This work was supported by the Australian Research Council (ARC) through grants CE170100012, DP220102518, DP220100178, DP190101506 and FT200100073. A.J.H. and I.O.R. are supported by an Australian Government Research Training Program Scholarship. S.C.S gratefully acknowledges the support of an Ernst and Grace Matthaei scholarship. B.R. acknowledges funding from the Royal Society University Research Fellowship URF/R1/180127, European Union's H2020 Framework Programme/European Research Council Starting Grant (852674  AngstroCAP), Philip Leverhulme Prize PLP-2021-262, EPSRC strategic equipment grant (EP/W006502/1). R.Q. acknowledges CSC scholarship. We acknowledge G. Whitehead for the XRD measurements, and support from E.~Mcinnes and F.~Tuna for the operation and data collection from EPSRC EPR National Research Facility's SQUID magnetometer supported by EP/S033181/1.

\end{acknowledgments}


\appendix

\section{Sample fabrication}
\label{apx_stru_char}

\begin{figure}
    \centering
    \includegraphics[width=0.45\textwidth]{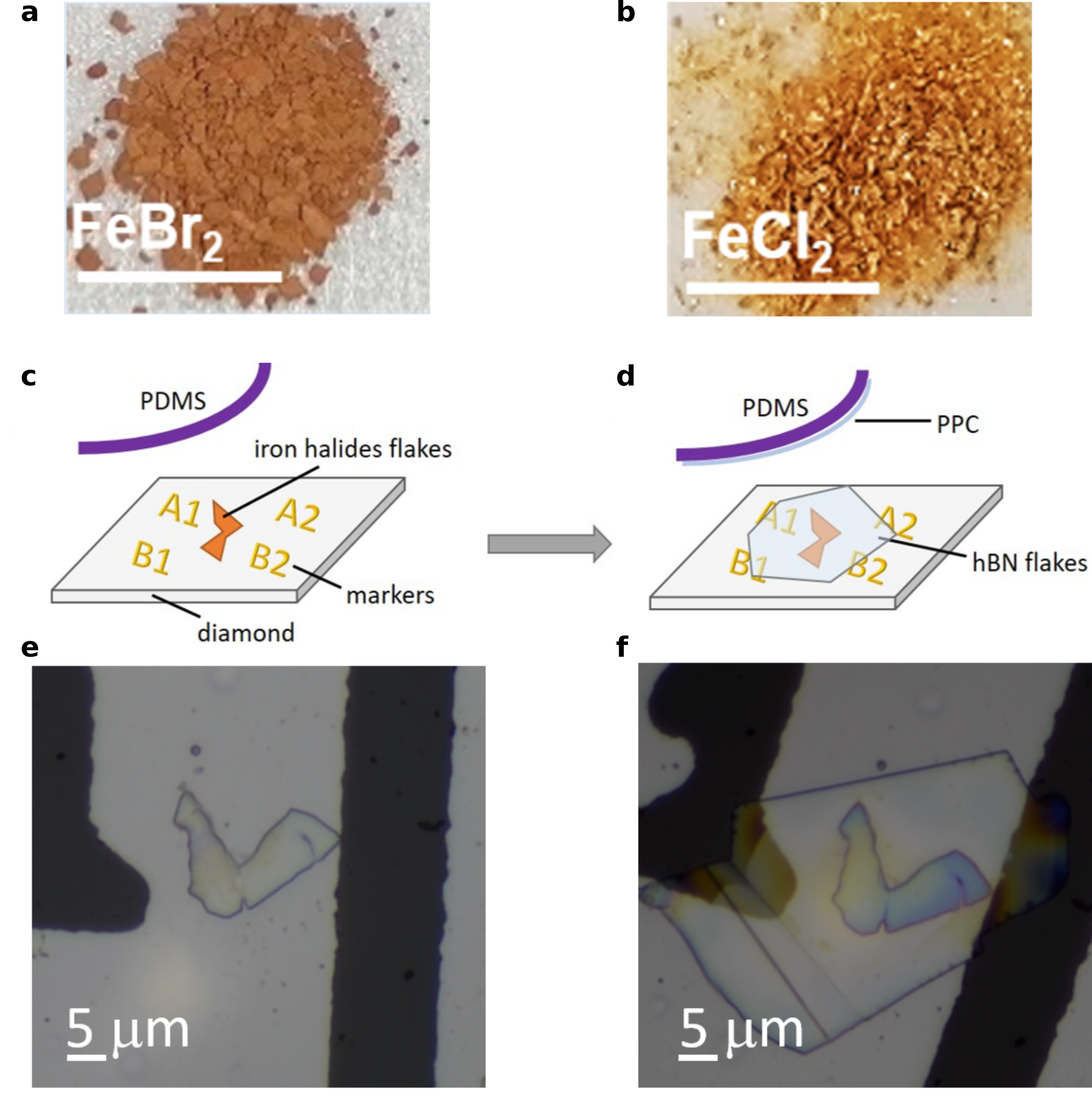}
    \caption{
    \textbf{(a)}~FeBr$_2$ and \textbf{(b)}~FeCl$_2$ bulk powders used for direct mechanical exfoliation, each scale bar is 5~mm long. \textbf{(c,d)}~Schematic sealing process in which thin iron halides flakes are covered by a protective hBN flake (thickness $<$40~nm). \textbf{(e,f)}~Optical images for a FeBr$_2$ flake at the \textbf{c,d}~stages, respectively.
    }
    \label{fig_apx_transfer}
\end{figure}

As described in the main text, FeBr$_2$ and FeCl$_2$ thin flakes were obtained from mechanical exfoliation of commercially sourced powders, which are displayed in \textcolor{blue}{Figure~\ref{fig_apx_transfer}(a,b)}. All exfoliation and transferring steps were carried out in a glovebox at room temperature. For the dry transfer, we first prepared a PDMS sheet using a PDMS elastomer kit composed of two liquid components (component A: the base; component B: the curing agent) that were mixed to a ratio of 10:1~\%w. A 2~$\times$~2~mm$^2$ square-shape PDMS sheet was then placed on a glass slide, which was previously treated by oxygen plasma for 10~min. The diamond substrate was also pre-treated with plasma during 2~min to ensure strong adhesion for the samples. After that, the iron halides flakes were mechanically exfoliated onto the PDMS sheets, then transferred directly onto a specified region of a diamond substrate with the transfer angle remaining horizontal, and finally heated at 30$-$40~°C for 5~min. In order to protect the samples from ambient degradation, a PDMS slice was first treated with plasma for 2~min and then coated with polypropylene carbonate (PPC). This PDMS/PCC stack was used to pick a hexagonal boron nitride (hBN) flake (thickness $<$40~nm) and place it onto the iron halides flakes at 60~°C, effectively sealing them. \textcolor{blue}{Figure~\ref{fig_apx_transfer}(c-f)} shows a scheme of the sealing process along with optical images of a FeBr$_2$ flake before and after being covered.

\begin{figure}
    \centering
    \includegraphics[width=0.5\textwidth]{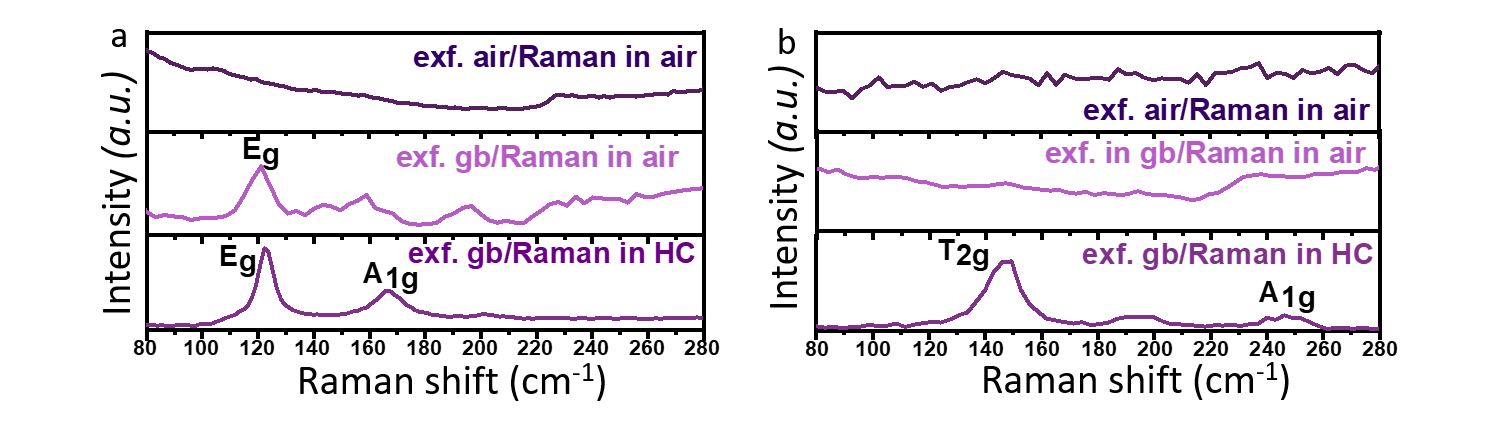}
    \caption{
    \textbf{(a)}~FeBr$_2$ and \textbf{(b)}~FeCl$_2$ thin flakes' stability studied by Raman spectroscopy under different preparation and evaluation conditions: both exfoliation and Raman control in air (top panels), exfoliation in glovebox and Raman control in air (middle panels) and exfoliation in glovebox and Raman control in hermetic cell (bottom panels). Characteristic vibrational modes of FeBr$_2$ ($E_{\rm g}$ and $A_{\rm 1g}$) and FeCl$_2$ ($T_{\rm 2g}$ and $A_{\rm 1g}$) structures are clearly present in the most protected conditions, and they gradually disappear when samples are exposed to air.
    }
    \label{fig_apx_stability}
\end{figure}

The evaluation of the flakes' stability was assessed by three different preparation/control methods: 1. exfoliation in air and posterior Raman spectroscopy in air; 2. exfoliation in inert environment (glovebox full of Ar atmosphere) and Raman in air; 3. exfoliation in inert environment and Raman in a hermetic cell. Raman spectra are shown in \textcolor{blue}{Figure~\ref{fig_apx_stability}}, where characteristic vibrational modes of FeBr$_2$ ($E_{\rm g}$ and $A_{\rm 1g}$)~\cite{Lockwood1978} and FeCl$_2$ ($T_{\rm 2g}$ and $A_{\rm 1g}$)~\cite{Johnstone1978TemperatureScattering} can be seen for the best protected case (method 3), then a slight degradation when exposing the samples to air for the Raman control (method 2), and finally losing the iron halides modes when samples are not protected at all (method 1).

\begin{figure}
    \centering
    \includegraphics[width=0.5\textwidth]{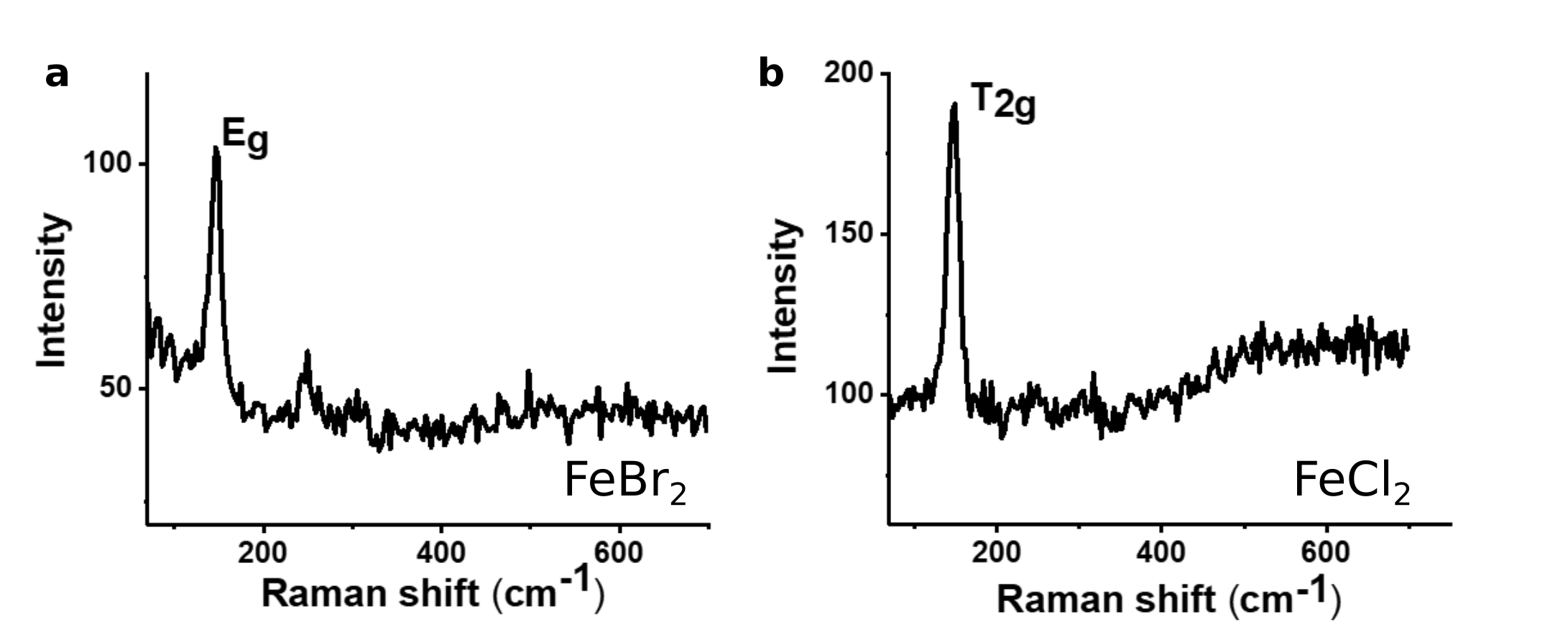}
    \caption{
    Raman evaluation for main samples \textbf{(a)}~FeBr$_2$ and \textbf{(b)}~FeCl$_2$. The main vibrational modes $E_{\rm g}$ of FeBr$_2$ and $T_{\rm 2g}$ of FeCl$_2$ are clearly present and provide evidence that the samples were not degraded.
    }
    \label{fig_apx_Raman}
\end{figure}

Samples FeBr$_2$ and FeCl$_2$ from the main text were exfoliated in glovebox but we didn't immediately evaluate the Raman results to ensure that the flakes were not degraded. Only after all the magnetic measurements were done we proceeded to the Raman evaluation, as shown in \textcolor{blue}{Figure~\ref{fig_apx_Raman}(a,b)} for FeBr$_2$ and FeCl$_2$ samples, respectively, confirming that the flakes were not degraded and that they exhibit the characteristic iron vibrational modes.

\begin{figure}
    \centering
    \includegraphics[width=0.5\textwidth]{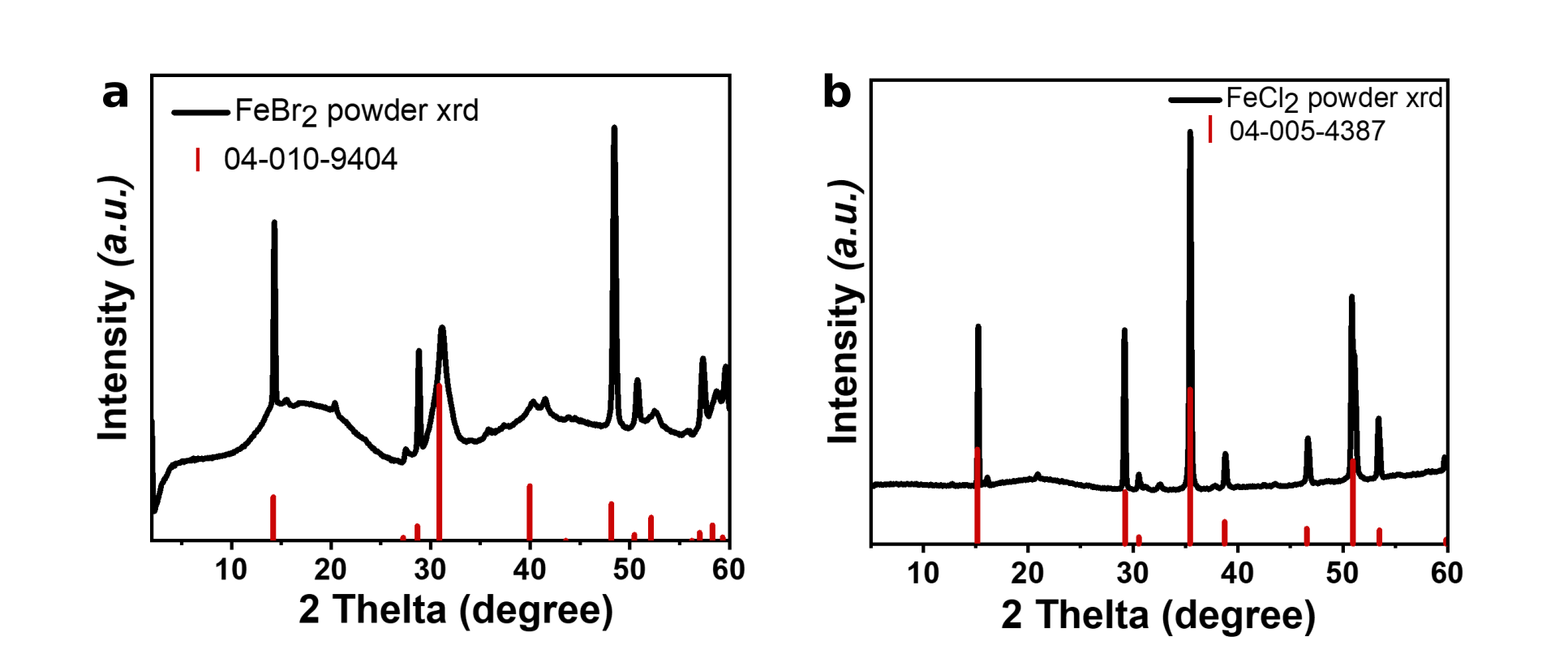}
    \caption{
    X-Ray Diffraction for bulk \textbf{(a)}~FeBr$_2$ and \textbf{(b)}~FeCl$_2$ materials before mechanical exfoliation. The experimental patterns (black) are compared with peak positions of reference powder patterns (red vertical lines) to show that they share the same crystal structure ($\rm P\overline{3}m1$). 
    }
    \label{fig_apx_XRD}
\end{figure}

The crystallographic structure was studied by a Rigaku FR-X DW diffractometer on the bulk materials before mechanical exfoliation (signals coming from the thin flakes are too weak for this study), see \textcolor{blue}{Figure~\ref{fig_apx_XRD}(a,b)} for FeBr$_2$ and FeCl$_2$, respectively. The experimental signals (black curves) are compared with peak positions of reference powder patterns (red vertical lines) \#04-010-9404 for FeBr$_2$~\cite{Ito1999} and \#04-005-4387 for FeCl$_2$~\cite{Rozenberg2009Pressure-inducedMeasurements}, evidencing that they share the same $\rm P\overline{3}m1$ crystal structures. 

\section{NV widefield setup}
\label{apx_NV_widefield}

The stray fields sensed by the NV spin defects were measured by optically detected magnetic resonance (ODMR) by sweeping the MW frequency and looking for photoluminescence (PL) contrast, which is indicative of a spin transition related to the stray fields intensity. A further explanation can be found in \cite{Scholten2022ImagingMicroscope}. The diamond sensors were made from type Ib high-pressure high-temperature (HPHT) diamond substrates, purchased from Element Six, with a N density of $\approx 50$~ppm. Dense NV layers ($<5$~ppm) were created through ion implantation: 2~MeV Sb implantation at a dose of $2 \times 10^{11}$~cm$^{-2}$ for the initial set of measurements, and 100~keV C implantation at a dose of $1\times 10^{13}$~cm$^{-2}$ for the hydrated sample measurements. These implantation procedures result in NV sensing layers of  $\approx 500$~nm (2~MeV Sb) and $\approx 200$~nm (100~keV C) thickness, which sets the upper bound of the sample-sensor standoff in our experiments when taken in combination with the thickness of the Al/Al$_2$O$_3$ coating $\approx 150$~nm. The surface orientation of the diamond sensors was $\langle$111$\rangle$, except for the hydrated sample measurements which used a $\langle 100 \rangle$ diamond.

The MW delivered by the gold resonator was produced by a signal generator Rohde \& Schwarz SMB100A, gated by a switch Mini-Circuits ZASWA-2-50DR+ and amplified by Mini-Circuits HPA-50W-63. A 532~nm laser (Laser Quantum Ventus) and an acousto-optic modulator (AA~Opto-Electronic MQ180-A025-VIS) were used to initialize and readout the NV layer, synchronized with the sCMOS camera PL acquisition and a SpinCore PulseBlaster ESR-PRO 500 MHz card.

 
\section{Energy density for the single domain model}
\label{apx_Edens}

Our single domain model considers an AF flake with $N$~layers as a two-sublattice system, defined by odd an even layers, each of them having their own magnetization ($\vec{M}_{\rm odd}$ and $\vec{M}_{\rm even}$). The orientation for each sublattice magnetization as a function of the bias field can be obtained by minimizing the energy density proposed in \textcolor{blue}{Equation~\ref{eq_E}}. Assuming $c$-axial cylindrical symmetry, before reducing the system to two sublattices and considering each layer~$i$ independent, the energy density terms can be described as follows:
$$ E_{\rm Zeeman} = - M_{\rm S} \left[\frac{1}{N}\sum_{i}^{N} (\vec{e_i} \cdot \vec{B_0}) \right] \, , $$
$$ E_{\rm demag} = \frac{1}{2}\mu_0 M_{\rm S}^2 \left[\frac{1}{N}\sum_{i}^{N} \cos(\theta_{i}) \right]^2 \, ,$$
$$ E_{\rm magcrys} = K_{\rm u} \left[\frac{1}{N}\sum_{i}^{N} \sin^2(\theta_{i}) \right] \, ,$$
$$ E_{\rm AF} = J_{\rm AF} \left[ \frac{1}{N}\sum_{i}^{N-1} \cos(\theta_{i}-\theta_{i+1}) \right] \, ,$$

\noindent where $M_{\rm S}$ is the saturation magnetization, $\vec{e_i} = \vec{M_{i}}/|\vec{M_{i}}|$ is the magnetization orientation for the layer $i$, $\theta_{i}$ is the polar angle for $\vec{e_i}$, $K_{\rm u}$ is the axial magnetocrystalline anisotropy and $J_{\rm AF}$ is the AF exchange constant (absolute value). Each term will compete to align each layer's magnetization in different directions: $E_{\rm Zeeman}$ looks for alignment along the bias field direction, $E_{\rm demag}$ tries to keep them in the basal plane while $E_{\rm magcrys}$ in the $c$~axis, and $E_{\rm AF}$ favors anti-parallel magnetic orientation between adjacent layers $i$ and $i+1$. 

If we consider that odd and even layers behave as sublattices with $N_{\rm odd}$ and $N_{\rm even}$ number of layers, respectively, as expected for type-A antiferromagnetism, then the $N$ free variables $\{\vec{M_{i}}\}$ are simplified to $\vec{M}_{\rm odd}$ and $\vec{M}_{\rm even}$. Furthermore, setting the bias field direction along the $c$~axis, the previous energy terms reduce to those stated in the main text.

Numerical values for the energy parameters were chosen according to literature values and checking that our model reproduces the expected bias breaking field $B_{\rm 0,b}$ for which the spin-flip transition occurs and the magnetization aligns (almost) completely with $\vec{B}_0$ when it is applied along the $c$~axis, see \textcolor{blue}{Table~\ref{table_Eparams}}. All values are given for temperature $T=4$~K.

\begin{table}
    \caption{Numerical values for energy density equation at 4~K.}
    \label{table_Eparams}
    \resizebox{\columnwidth}{!}{%
    \begin{tabular}{ |p{3cm}|p{3cm}|p{3cm}|p{3cm}|p{3cm}|  }
     \hline 
     Parameter & FeBr$_2$ & FeCl$_2$ & References \\
     \hline
     $M_{\rm S}$ [MA/m]         & 0.49          & 0.61          
     & \cite{Pelloth1995,Jacobs1967MagneticFeCl2,Jacobs1964} \\
     $K_{\rm u}$ [kJ/m$^3$]     & 700           & 400           
     & \cite{Botana2019,Birgeneau1972} \\
     $J_{\rm AF}$ [kJ/m$^3$]    & 650           & 200           
     & \cite{Pleimling1997,Balucani1985,Fert1973} \\
     $B_{\rm 0,b}$ [T]          & $\approx$3.1 & $\approx$1.1  
     & \cite{Pelloth1995,Jacobs1967MagneticFeCl2,Jacobs1964} \\
     \hline
    \end{tabular}
    }
\end{table}


\section{Hysteresis measurements}
\label{apx_NoHyst}

\begin{figure}
    \centering
    \includegraphics[width=0.45\textwidth]{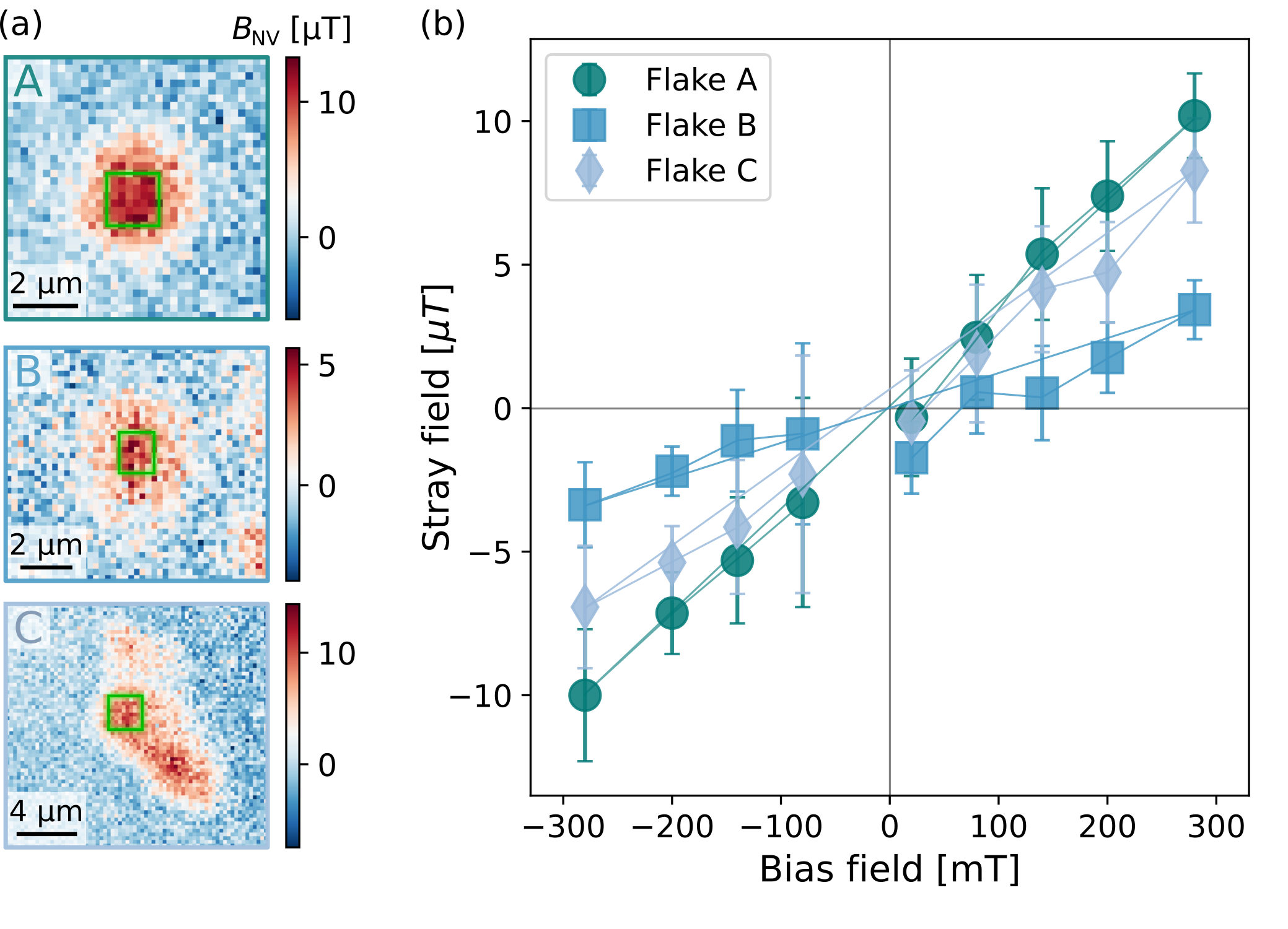}
    \caption{
    Stray field hysteresis measurements for different flakes. \textbf{(a)}~Stray field maps at 280~mT bias fields for three flakes named A, B, C (from top to bottom) in which the stray fields are averaged within the green delimited region. \textbf{(b)}~Stray vs bias fields, cycling the bias field from zero to the maximum positive value, then to the maximum negative value and finally back to zero. All flakes show negligible magnetization and coercivity.
    }
    \label{fig_apx_NoHyst}
\end{figure}

When performing magnetic measurements on the samples described in the main text, no hysteresis was observed after cycling the bias fields. We explicitly investigated this on thinner FeCl$_2$ flakes (estimated to be 50~nm thick), where it may be expected that hard FM behaviors in uncompensated layers (or at least diverse behavior compared to the bulk) could be more apparent~\cite{Tan2018HardFe3GeTe2}. \textcolor{blue}{Figure~\ref{fig_apx_NoHyst}} shows the results for three different samples, averaging the stray fields in a delimited region (error bars represent one standard deviation). In all cases, we observe a linear trend consistent with the results in the main text, showing negligible remanent magnetization and coercivity.

\section{Paramagnetic model}
\label{apx_PM_model}

Paramagnetic theory~\cite{Kittel1957IntroductionPhysics} allows to build a model in which the whole flake is considered PM, then each Fe$^{2+}$ ion contributes to the total magnetization $M_{\rm PM}$ according to the quantum numbers $S$,$L$,$J$ (see \textcolor{blue}{Equation~\ref{eq_PM}} on the main text). The Land\'e factor~$g$ is described by:
$$ g = 1 + \frac{J(J+1)+S(S+1)-L(L+1)}{2J(J+1)} \, ,$$

\noindent which gives $g=1.5$ for a free ion ($J=4$, $S=L=2$) \cite{Ropka2001} and $g=2$ for both the special case of a free ion with quenched angular momentum ($J=S=2$, $L=0$)~\cite{Niemeyer2012SpinClusters} and a bounded ion ($J=S=1$, $L=0$)~\cite{Pelloth1995}. This constant scales the magnetization~$M_{\rm PM}$, while the functional form can be described by the Brillouin function:
$$ B_{\rm J}(x) = \frac{2J+1}{2J} \coth \left( \frac{(2J+1)}{2J}x \right)-\frac{1}{2J} \coth \left( \frac{x}{2J} \right) \, ,$$

\noindent where the argument $x = g J \mu_{\rm B} B_0 / k_{\rm B} T$ includes the bias field~$B_0$ and temperature~$T$ dependence.

\begin{figure}
    \centering
    \includegraphics[width=0.45\textwidth]{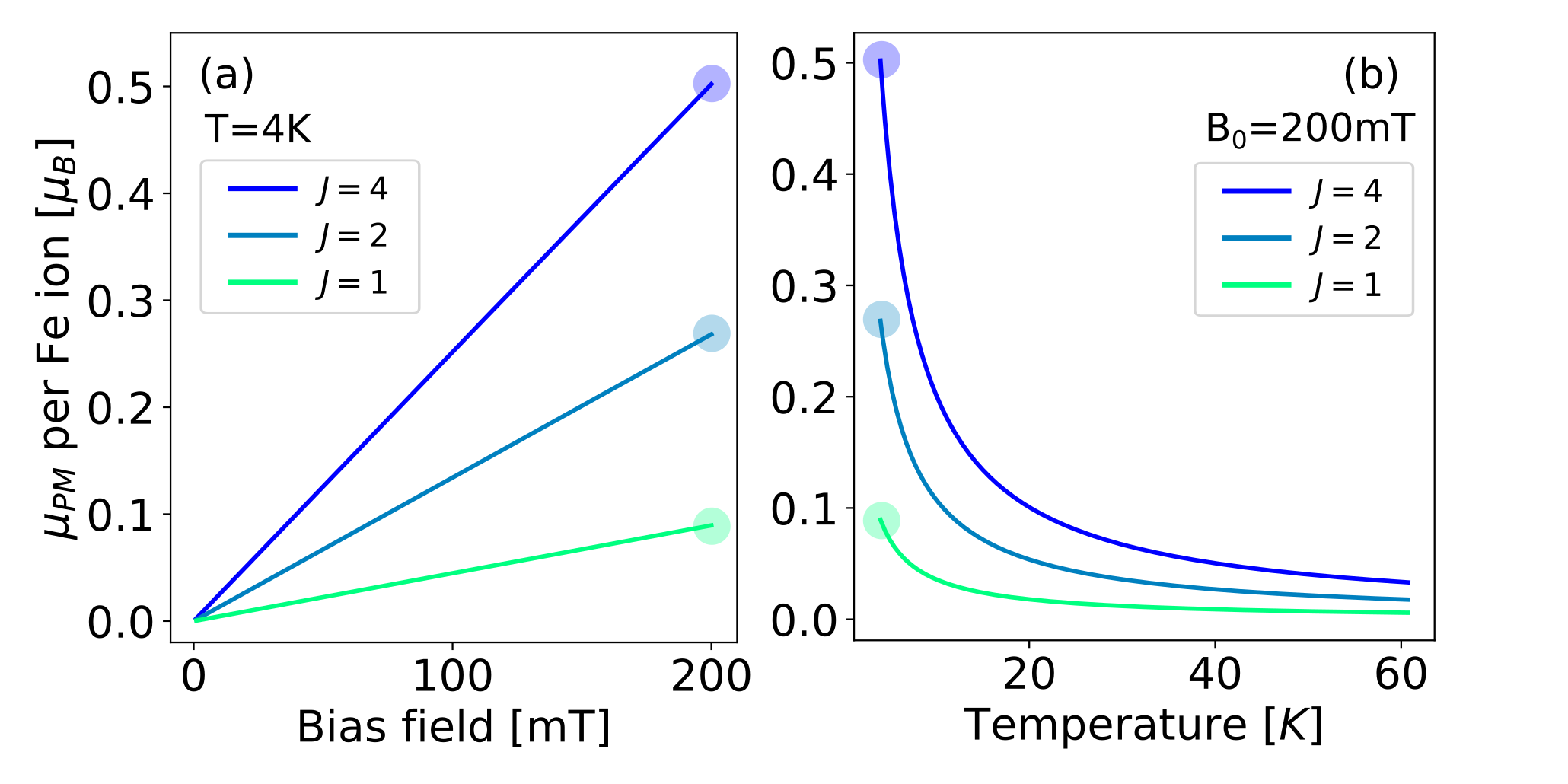}
    \caption{
    Calculated paramagnetic moment per Fe ion (in Bohr magneton units) as a function of: \textbf{(a)}~bias field at 4~K and \textbf{(b)}~temperature at 200~mT. The values at 4~K and 200~mT have been identified with shaded circles in both plots.
    }
    \label{fig_apx_muPM}
\end{figure}

\textcolor{blue}{Figure~\ref{fig_apx_muPM}(a,b)} shows the PM moment per Fe ion $\mu_{\rm PM}=M_{\rm PM}/n$ ($n$ is the number of atoms per unit volume) as a function of bias field at $T=4$~K and as a function of temperature at $B_0=200$~mT, respectively. The three quantum models (labeled by the quantum number~$J$) have been included, confirming that the bounded ion $J=1$ exhibits the weakest magnetic response. Notice that these curves are common for both FeBr$_2$ and FeCl$_2$ lattices, since we are neglecting the contribution of Br or Cl ions.

The magnetization curves $M_{\rm PM}$ can be finally calculated using the values $n=13.4$~nm$^{-3}$ for FeBr$_2$~\cite{Yelon1975} and $n=16.2$~nm$^{-3}$ for FeCl$_2$~\cite{Pasternak1976}.


\section{Temperature dependence expectations for bulk antiferromagnetism}
\label{apx_AF_MvsT_expec}

The behavior of volume magnetic susceptibilities $\chi_{\rm v}$ of AF materials as a function of temperature $T$ are well studied and described, see \cite{Mugiraneza2022Tutorial:Law, Kittel1957IntroductionPhysics}. Below the N\'eel temperature~$T_{\rm N}$, the susceptibility should decay from its maximum value $\chi_{\rm v}(T_{\rm N})$ towards a remanent susceptibility $\chi_{\rm v}(0\,\text{K})$ (ideally zero), as demonstrated experimentally~\cite{Jacobs1964,Jacobs1967MagneticFeCl2,Alben1969}. Above $T_{\rm N}$, the susceptibility should decay similar to a paramagnet (as $1/T$) but modified by the Weiss temperature $\theta_{\rm W}$, then $\chi_{\rm v}(T>T_{\rm N}) \propto 1/(T+\theta_{\rm W})$. For bulk FeCl$_2$ it is reported that $\theta_{\rm W}=48$~K~\cite{Kittel1957IntroductionPhysics}, whereas for FeBr$_2$ we couldn't find any reference value. For a qualitative model, we estimate this parameter from the ratio between the N\'eel temperatures: 
$$ \theta_{\rm W}[\text{FeBr}_2] = \theta_{\rm W}[\text{FeCl}_2] \times 
\frac{T_{\rm N}[\text{FeBr}_2]}{T_{\rm N}[\text{FeCl}_2]} \approx 28K \, .$$


\section{Alternative NV axes orientation}
\label{apx_dia_NVprojections}

\begin{figure}
    \centering
    \includegraphics[width=0.45\textwidth]{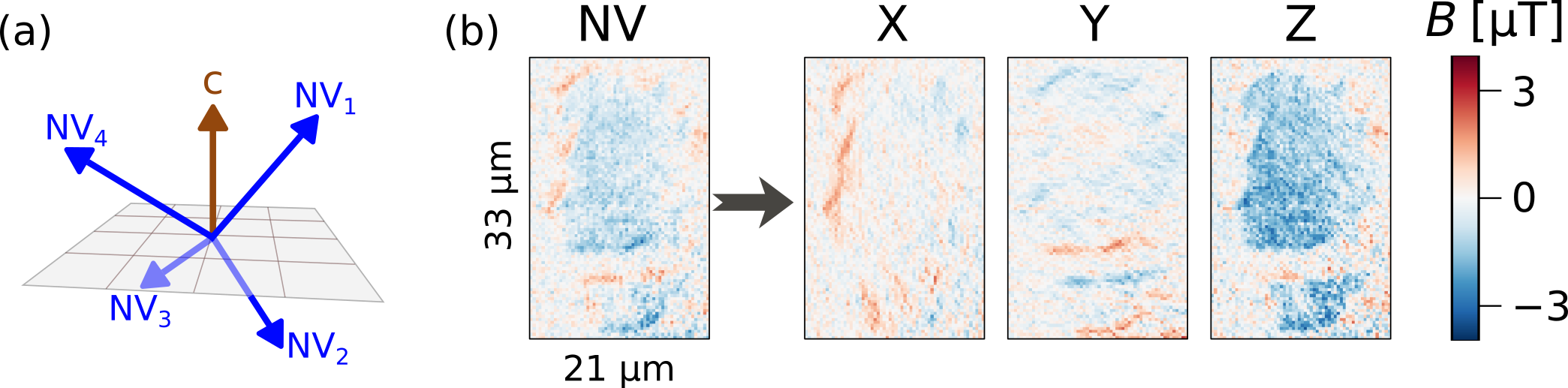}
    \caption{
    \textbf{(a)}~Orientation of the four NV axes relative to the crystal anisotropy $c$~axis.
    \textbf{(b)}~Full stray field vector reconstruction from the original map measured along the NV$_2$ axis at 280~mT bias field (left), projected along the (x,y,z) directions (right).
    }
    \label{apx_dia_NVaxes}
\end{figure}

\textcolor{blue}{Figure~\ref{apx_dia_NVaxes}(a)} shows the NV axes orientation relative to the crystal anisotropy $c$~axis, along the directions (1,1,1), (-1,1,-1), (1,-1,-1) and (-1,-1,1) in Cartesian coordinates. As the NV spin Hamiltonian is invariant under a change in sign of the bias field, the positive or negative prescription is arbitrary, and in the following we define the positive direction to be that which has a positive projection in the lab $z$~direction, coincidental with the $c$~axis. The stray fields detected are also defined in this way.

From a magnetic map on the thin flakes sample measured along one of the NV axes (with both parallel and perpendicular components relative to the c~axis), the full stray field vector $\vec{B}$ can be reconstructed, as explained in~\cite{Scholten2022ImagingMicroscope}. \textcolor{blue}{Figure~\ref{apx_dia_NVaxes}(b)} shows an example of this process, in which the original stray field map (left) measured along the (-1,1,-1) NV$_2$ axis is used to obtain the (x,y,z) projections of $\vec{B}$.


\section{Simulations for diamagnetic sample}
\label{apx_dia_sims}

\begin{figure}
    \centering
    \includegraphics[width=0.45\textwidth]{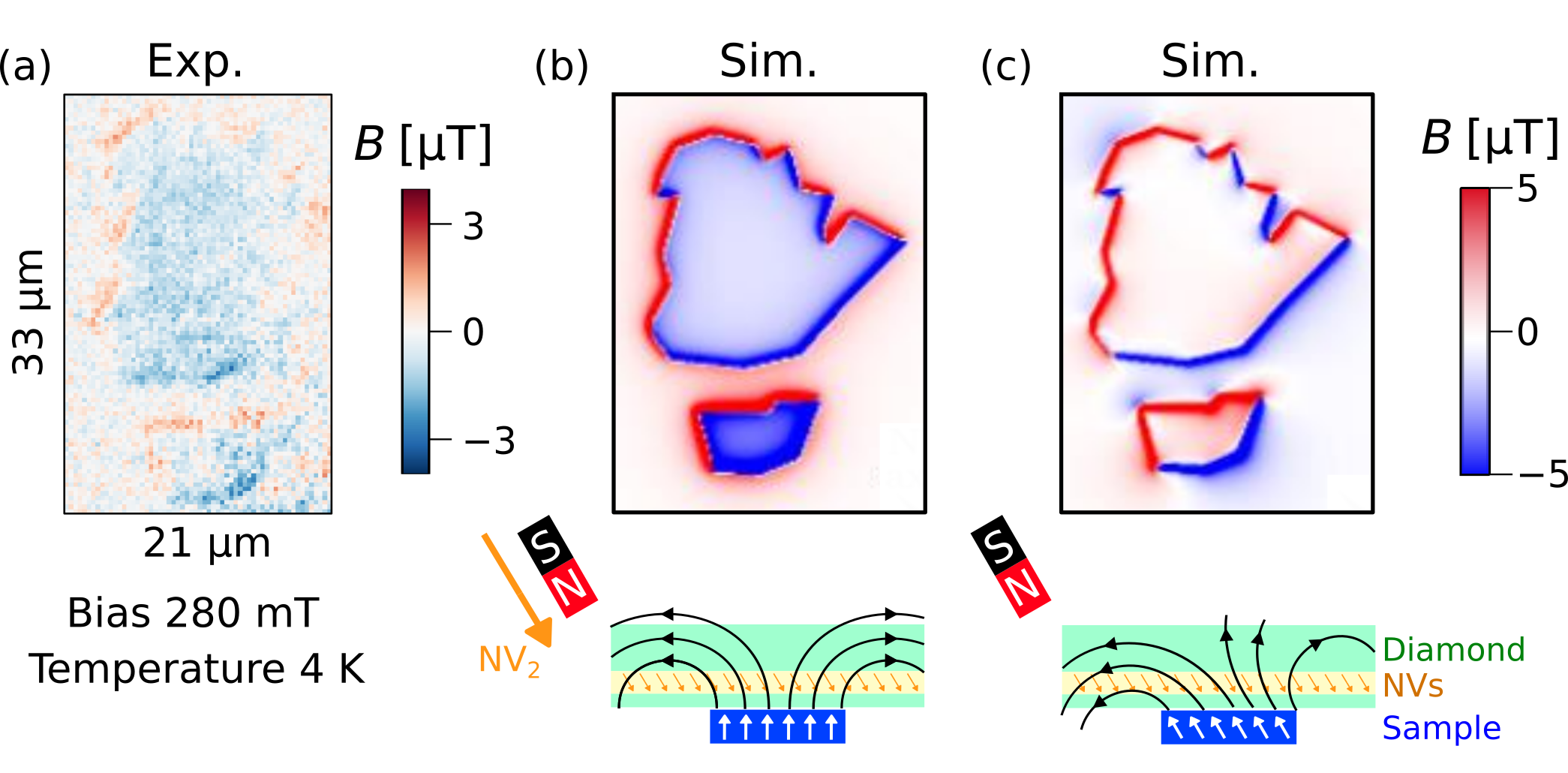}
    \caption{
    \textbf{(a)}~Experimental $c$-projection of the stray fields generated by a FeBr$_2$ flake under a 280~mT bias field applied along the NV$_2$ axis at 4~K, sensed by the NV centers at 500~nm standoff distance. 
    \textbf{(b,c)}~Stray field $c$-projections simulated under the same bias and temperature conditions assuming \textbf{(b)}~a strong anisotropy $c$~axis and \textbf{(c)}~no anisotropy axis.
    }
    \label{fig_apx_dia_sim}
\end{figure}

The diamagnetic response of a thin FeBr$_2$ flake at 4~K and under a 280~mT bias field applied along the NV$_2$ axis was simulated and compared to the experimental measurements, displayed in \textcolor{blue}{Figure~\ref{fig_apx_dia_sim}(a)}. The simulations describe the stray fields according to two different magnetic anisotropy models: strong anisotropy along the sample's $c$~axis and no magnetic anisotropy (\textcolor{blue}{Figure~\ref{fig_apx_dia_sim}(b,c)}, respectively). In the former case, the flake's magnetic moments try to align in the opposite direction to the bias field, but the crystalline anisotropy prevents them to align perfectly and pins them to the $c$~axis. In the later case, there are no restrictions for the magnetic moments and they align perfectly opposite to the bias field. When the stray fields are projected along the $c$~axis, the strong anisotropy model shows a better agreement with the experimental results, which is evidenced by the non-zero stray fields matching the inner regions of the flake, which should vanish in the no-anisotropy case.

\bibliography{references}

\end{document}